\documentclass[journal]{IEEEtran}

\usepackage{graphicx}
\usepackage{cite}

\begin{document}

\newcommand{\cm}[1]{\ensuremath{ {{\rm cm}^{#1}}}}
\newcommand{\ntp}[2]{\ensuremath{#1\times10^{#2} } }
\newcommand{\asb}[2]{\ensuremath{#1_{\rm #2} }}

\begin{titlepage}

\thispagestyle{empty}
\def\thefootnote{\fnsymbol{footnote}}       

\begin{center}
\mbox{ }

\end{center}
\begin{center}
\vskip 1.0cm
{\Huge\bf
Modeling of Charge Transfer Inefficiency in a 
}
\vspace{2mm}

{\Huge\bf
CCD~with~High-Speed~Column\,Parallel\,Readout
}
\vskip 1cm
{\LARGE\bf
Andr\'e Sopczak$^1$,
Salim Aoulmit$^2$,
Khaled Bekhouche$^1$,  
Chris Bowdery$^1$,
Craig Buttar$^3$,
Chris Damerell$^4$, 
Dahmane Djendaoui$^2$, 
Lakhdar Dehimi$^2$, 
Tim Greenshaw$^5$, 
Michal Koziel$^1$,  
Dzmitry Maneuski$^3$, 
Andrei Nomerotski$^6$,
Konstantin Stefanov$^4$,
Tuomo Tikkanen$^5$,
Tim Woolliscroft$^5$, 
Steve Worm$^4$
\bigskip}

{\Large
$^1$Lancaster University, UK\\
$^2$Biskra University, Algeria\\
$^3$Glasgow University, UK\\
$^4$STFC Rutherford Appleton Laboratory, UK\\
$^5$Liverpool University, UK \\
\smallskip
$^6$Oxford University, UK
}

\smallskip

\vskip 2.cm
\centerline{\Large \bf Abstract}
\end{center}

\vskip 1.5cm
\hspace*{-0.5cm}
\begin{picture}(0.001,0.001)(0,0)
\put(,0){
\begin{minipage}{\textwidth}
\Large
\renewcommand{\baselinestretch} {1.2}
Charge Coupled Devices (CCDs) have been successfully used in several high energy physics 
experiments over the past two decades. Their high spatial resolution and thin sensitive 
layers make them an excellent tool for studying short-lived particles. The Linear Collider 
Flavour Identification (LCFI) collaboration is developing Column-Parallel CCDs (CPCCDs) 
for the vertex detector of a future Linear Collider. The CPCCDs can be 
read out many times faster than standard CCDs, significantly increasing their operating speed.
An Analytic Model has been developed for the determination of the charge transfer inefficiency (CTI)
of a CPCCD. The CTI values determined with the Analytic Model agree largely with those from a
full TCAD simulation. The Analytic Model allows efficient study of the variation of the CTI
on parameters like readout frequency, operating temperature and occupancy.

\normalsize
\vspace{1.5cm}
\begin{center}
{\sl \large
Presented on behalf of the LCFI Collaboration at the \\
IEEE 2008 Nuclear Science Symposium, Dresden, Germany, and at the \\
11th Topical Seminar on Innovative Particle and Radiation Detectors 
(IPRD08) 2008, Siena, Italy
\vspace{-6cm}
}
\end{center}
\end{minipage}
}
\end{picture}
\vfill

\end{titlepage}

\newpage
\thispagestyle{empty}
\mbox{ }
\newpage
\setcounter{page}{0}

\title{Modeling of Charge Transfer Inefficiency in a CCD with
       High-Speed Column Parallel Readout}

\author{Andr\'e~Sopczak,~\IEEEmembership{Member,~IEEE,}\thanks{A. Sopczak is with Lancaster University, UK. 
Presented on behalf of the}\thanks{~~LCFI Collaboration; E-mail: andre.sopczak@cern.ch}
Salim~Aoulmit,\thanks{S. Aoulmit is with LMSM Laboratory Biskra University, Algeria}
Khaled~Bekhouche,\thanks{K. Bekhouche is with Lancaster University, UK}
Chris~Bowdery,\thanks{C. Bowdery is with Lancaster University, UK}
Craig~Buttar,\thanks{C. Buttar is with Glasgow University, UK}
Chris~Damerell,\thanks{C. Damerell is with STFC Rutherford Appleton Laboratory, UK}
Dahmane~Djendaoui,\thanks{D. Djendaoui is with LMSM Laboratory Biskra University, Algeria}
Lakhdar~Dehimi,\thanks{L. Dehimi is with LMSM Laboratory Biskra University, Algeria}
Tim~Greenshaw,\thanks{T. Greenshaw is with Liverpool University, UK}
Michal~Koziel,\thanks{M. Koziel is with Lancaster University, UK}
Dzmitry~Maneuski,\thanks{D. Maneuski is with Glasgow University, UK}
Andrei~Nomerotski,\thanks{A. Nomerotski is with Oxford University, UK}
Konstantin~Stefanov,\thanks{K. Stefanov is with STFC Rutherford Appleton Laboratory, UK}
Tuomo~Tikkanen,\thanks{T. Tikkanen is with Liverpool University, UK}
Tim Woolliscroft,\thanks{T. Woolliscroft is with Liverpool University, UK}
Steve Worm\thanks{S. Worm is with STFC Rutherford Appleton Laboratory, UK}}
\maketitle
\thispagestyle{empty}

\begin{abstract}          
Charge Coupled Devices (CCDs) have been successfully used in several high energy physics 
experiments over the past two decades. Their high spatial resolution and thin sensitive 
layers make them an excellent tool for studying short-lived particles. The Linear Collider 
Flavour Identification (LCFI) collaboration is developing Column-Parallel CCDs (CPCCDs) 
for the vertex detector of a future Linear Collider. The CPCCDs can be 
read out many times faster than standard CCDs, significantly increasing their operating speed.
An Analytic Model has been developed for the determination of the charge transfer inefficiency (CTI)
of a CPCCD. The CTI values determined with the Analytic Model agree largely with those from a
full TCAD simulation. The Analytic Model allows efficient study of the variation of the CTI
on parameters like readout frequency, operating temperature and occupancy.
\end{abstract}

\section{Introduction}
\begin{figure}
\includegraphics[width=\columnwidth,clip]{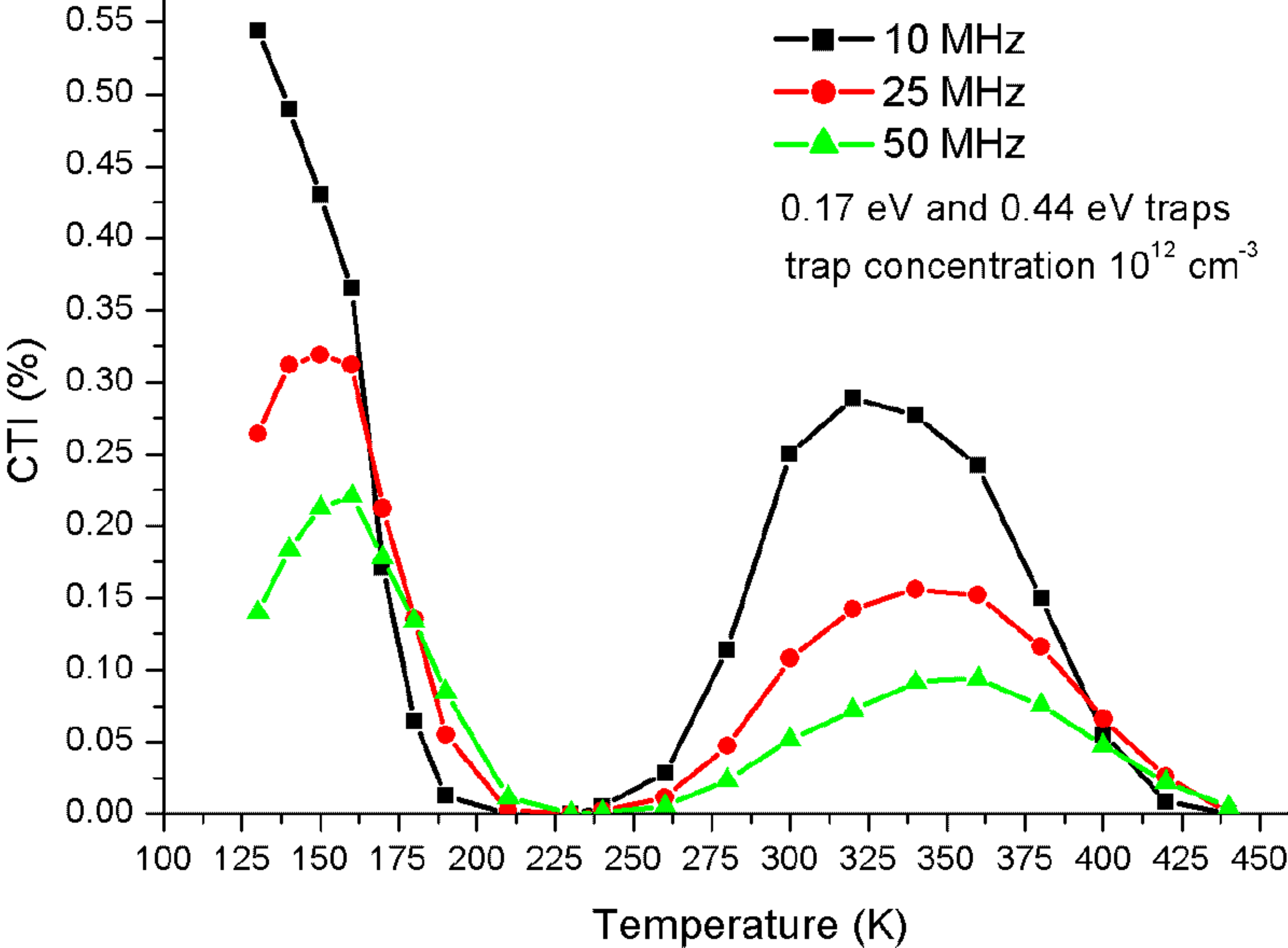}
\vspace*{-6mm}
\caption{CTI determined with TCAD simulations as a function of temperature for a two-phase CPCCD for two traps, 0.17 eV and 0.44 eV, 
with a concentration of $10^{12}$ cm$^{-3}$ and $1 \%$ hit (pixel) occupancy at readout frequencies 10, 25 and 50 MHz.}
\label{fig:ctivstemp_TCAD}
\end{figure}

Charge transfer inefficiency (CTI) is an important aspect in the CCD development for
operation in High Energy Physics colliders~\cite{demerell,stefanov,web}. The LCFI collaboration has been developing  
new CCD chips and testing them for about 10 years~\cite{demerell,stefanov,web,worm,greenshew}. 
Recently the focus of the simulations has been on CCDs with column parallel readout (CPCCD). 
Full TCAD~\cite{ISE} simulations for a CPCCD were performed for different readout frequencies and operating 
temperatures~\cite{Como2007,Hawaii,jinst}. An example of the CTI temperature dependence is shown in 
Fig.~\ref{fig:ctivstemp_TCAD} (from~\cite{jinst}). Full TCAD simulations are very CPU intensive. 
This has already been noted for the CCD simulations with a sequential readout~\cite{TNS07,IEEE06,IEEE05}. 
The CTI depends on many parameters, such as readout frequency and operating temperature. 
Some parameters are related to the trap characteristics like trap energy level, capture cross-section and 
trap concentration (density). Other factors are also relevant, such as the occupancy of the pixels (hits). 
It is well known that analytic charge transfer models can be used to study the CTI dependence on readout 
frequency and operating temperature~\cite{mohsan,hopkins,hardy}. For a comparison with full TCAD CTI simulation results, 
as shown in Fig.~\ref{fig:ctivstemp_TCAD}, we have developed Analytic Models for the CPCCD~\cite{Como2007,Hawaii,jinst}. 
The further development of these Analytic Models leads also to better understanding of the relevant parameters in order 
to reduce the CTI in future CPCCD prototypes. This paper addresses the inclusion of signal shape and clock voltage 
amplitude which leads to an improved Analytic Model for a CPCCD. The CTI obtained from the improved Analytic Model is compared 
with results from full TCAD simulations.
\vspace*{5mm}
\section{Analytic Model for CTI determination}
\begin{figure}
\centering
\includegraphics[width=\columnwidth,clip]{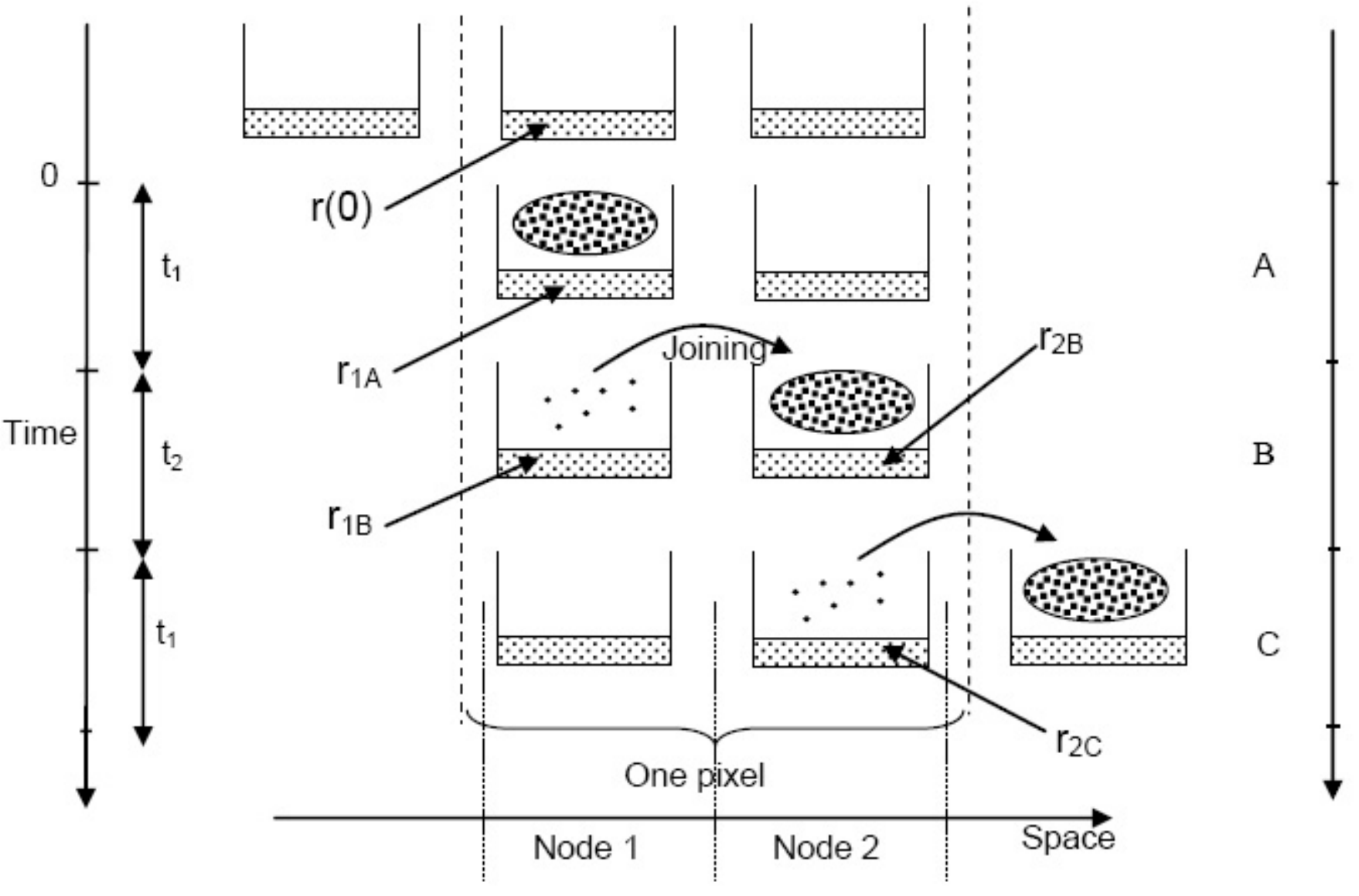}
\caption{Diagram of consecutive transfer of charge in a two-phase CCD. The diagram shows the charge transfer at different
 stages in time and space.}
\label{fig:diagram_transfer2}
\end{figure}

The Analytic Models~\cite{Como2007,Hawaii,jinst} describe 
the different steps in the charge transfer process and
the amount of the trapped charge with respect to the charge cloud in transfer. 
Figure~\ref{fig:diagram_transfer2} shows the consecutive charge transfer for a two-phase  
CPCCD in one pixel (2 nodes). Following the treatment by Kim~\cite{Kim}, based on earlier work by
Shockley, Read and Hall~\cite{Shockley} a defect at an energy below the bottom of the conduction
band is considered. Our model considers one single energy level and
includes the emission time $\tau_e$, and capture time $\tau_c$, in the differential equation  
\begin{equation}
\frac{dr}{dt}=\frac{1-r}{\tau_c}-\frac{r}{\tau_e}
\end{equation}
where $r$ is the fraction of filled traps. Initially the  fraction of filled traps is $r(0)$. At stage A the signal charge
packet arrives and  interacts with traps under node 1 during time $t_1$. This interaction leads
to the capture and emission process. By resolving the differential equation (1), the fraction of filled traps $r_{1A}$ under
node 1 during time $t_1$ (when the signal packet is present) is given by
\begin{equation}
r_{1A}(t_1) = [r(0)-\frac{\tau_s}{\tau_c}] \exp(-\frac{t_1}{\tau_s})+\frac{\tau_s}{\tau_c}
\end{equation} where $\tau_s=\tau_c\tau_e/(\tau_c+\tau_e)$.
 
At stage B charge moves to the next node and interacts with traps during time $t_2$ under this node. 
During this time electrons emitted from node 1 join the signal charge packet in the second node. 
Thus, the fraction of filled traps $r_{1B}$ under node 1 during time $t_1$ in the presence of the 
signal packet is given by
\begin{equation}
r_{1B}(t_2) = r_{1A}(t_1)\exp(-\frac{t_2}{\tau_e}). 
\end{equation}
At the same stage B, $r_{2B}$ is defined as the fraction of filled  traps  under node 2 during time $t_2$, thus,
\begin{equation}
r_{2B}(t_2) = [r(0)-\frac{\tau_s}{\tau_c}] \exp(-\frac{t_2}{\tau_s})+\frac{\tau_s}{\tau_c}.
\end{equation}

When the signal charge moves to the first node of the next pixel, stage C, electrons emitted during 
time $t_1$ can join the signal present at this node and the fraction of filled traps $r_{2C}$ 
under node 2 during time $t_1$ is given by 
\begin{equation}
r_{2C}(t_1) = r_{2B}(t_2)\exp(-\frac{t_1}{\tau_e}).
\end{equation}
The CTI is defined by the ratio of the charge loss under each node to the signal charge density $n_s$, thus,
\begin {equation}
CTI=\frac{N_t}{n_s}[r_{1B}(t_2)+r_{2C}(t_1)-2r(0)]
\end {equation}
where $N_t$ is the trap concentration, and $r(0)=\exp(-t_w/\tau_e)$ which is determined by considering the fact that initially all 
traps are filled and electrons are emitted during the waiting time $t_w$ between two signal charge packets. For the case $t_1=t_2=t$, 
the combination of the previous equations leads to
\begin{eqnarray}
CTI&=&2\frac{N_t}{n_s}[1-\exp(-t(\frac{1}{\tau_c}+\frac{2}{\tau_e}))]\times\nonumber\\
& &[(\frac{\tau_s}{\tau_e}\frac{(1-\exp(-\frac{t}{\tau_s}))}{(1-\exp(-t(\frac{1}{\tau_s}+\frac{1}{\tau_e})))})\exp(-\frac{t}{\tau_e})\nonumber\\
& &-\exp(-\frac{t_w}{\tau_e})].
\end{eqnarray}

\section{Analytic Model CTI results}
\begin{figure}
\centering
\includegraphics[width=\columnwidth,clip]{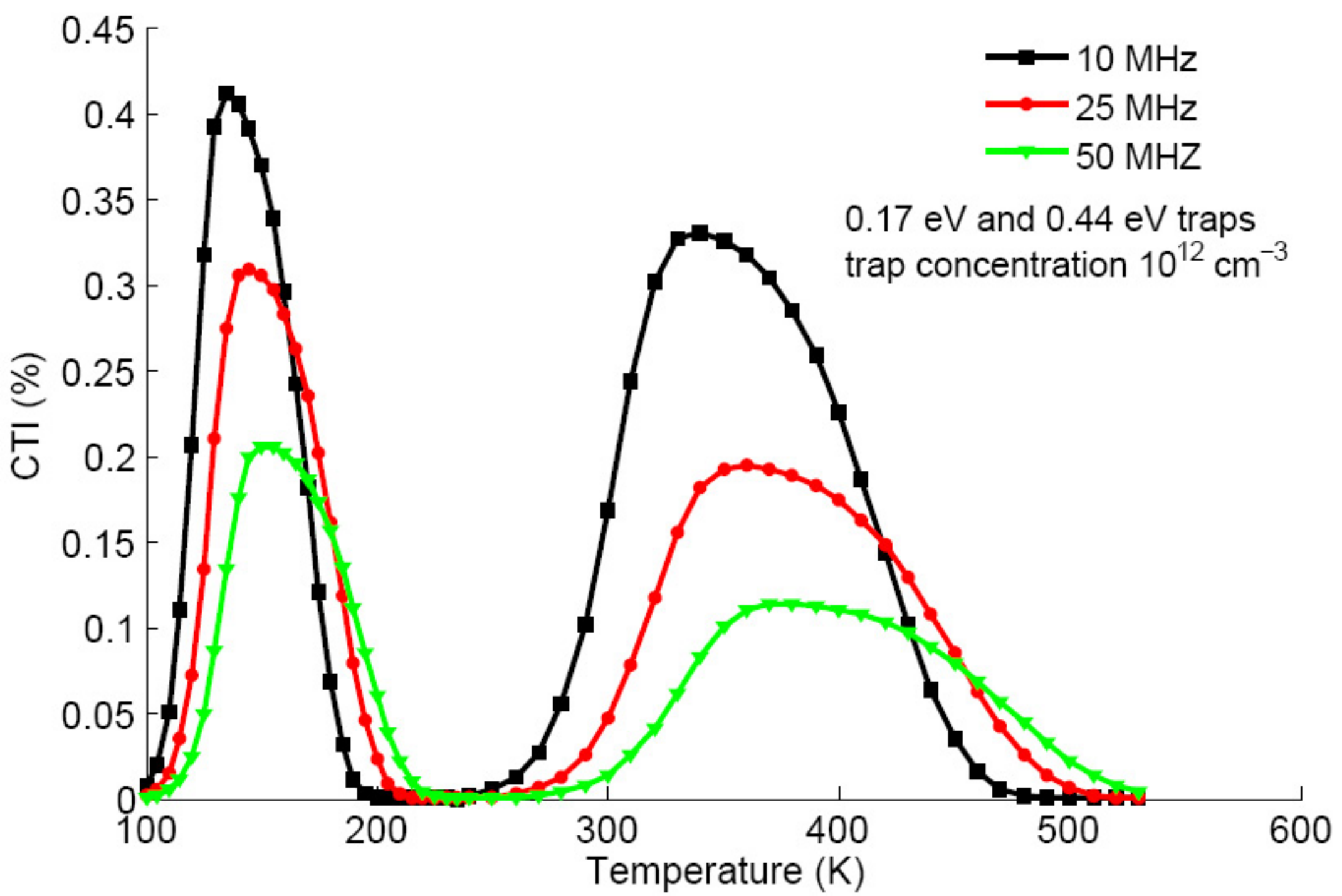}
\vspace*{-5mm}
\caption{CTI values from an Analytic Model as a function of temperature in a two-phase CPCCD 
for the two traps, 0.17 eV and 0.44 eV with a concentration of  $10^{12}$ cm$^{-3}$ and $1\%$ hit (pixel) 
occupancy at readout frequencies 10, 25 and 50 MHz.}
\label{fig:CTIvsTemp_1744_freq125b}
\end{figure}

The CTI dependence on readout frequency and operating temperature has been explored using an Analytic Model based on Eq.~(7). 
Figure~\ref{fig:CTIvsTemp_1744_freq125b} shows the CTI results from the Analytic Model at different frequencies for 
temperatures between 100\,K and 550\,K. The CTI increases as the readout frequency  decreases. For 
higher readout frequencies there is less time to trap the passing signal, thus the CTI is reduced.
At high temperatures the emission time is so short that the trapped charges can rejoin the passing signal.

\section {Comparison between full TCAD simulation and Analytic Model regarding signal shape effect}
\begin{figure}
\centering
\includegraphics[width=\columnwidth,clip]{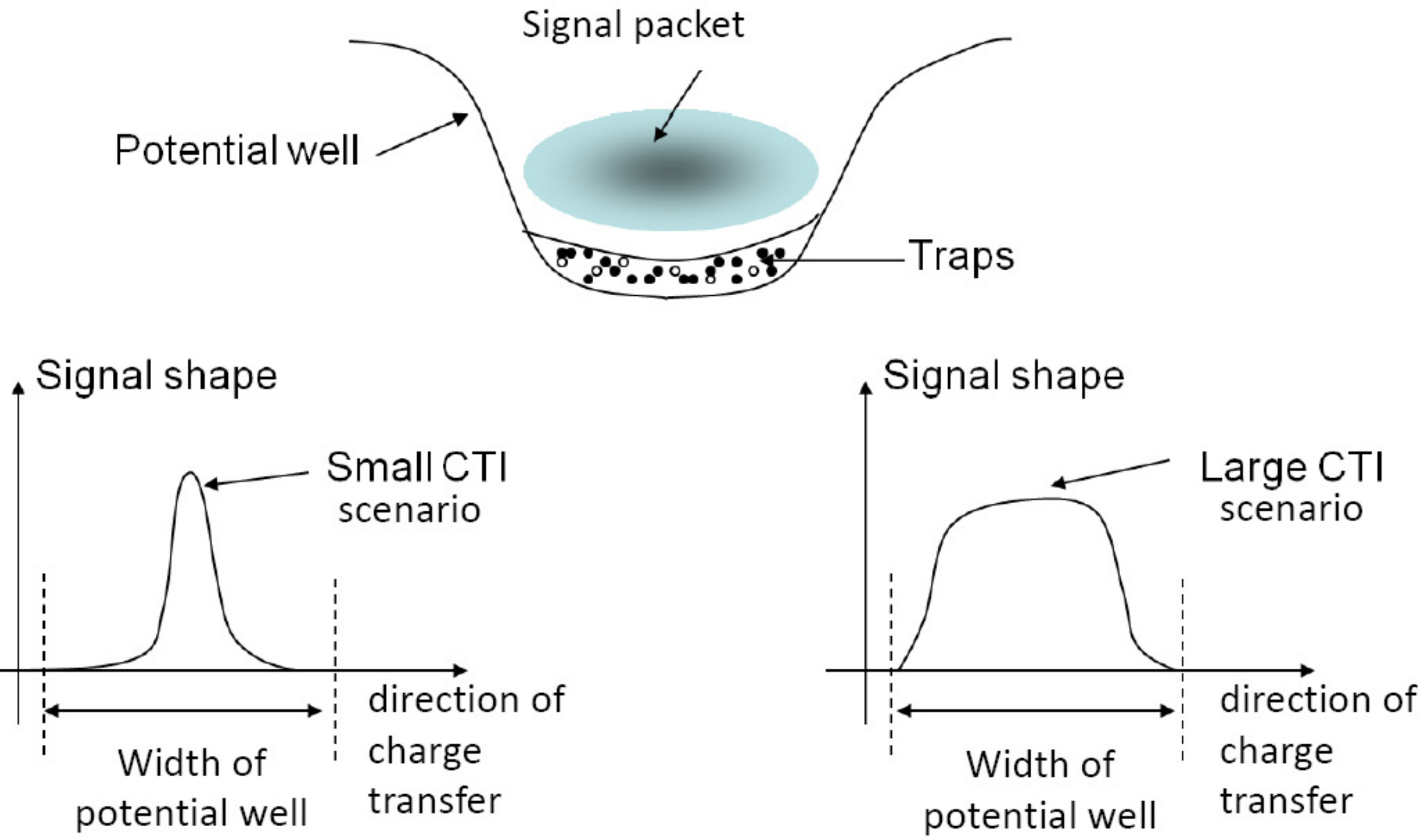}
\caption{Diagram of the signal packet in a potential well and effect of the signal shape
 on the expected CTI. Upper part: the density of a signal packet in a  potential well decreases gradually 
from the centre of the packet. Lower part: expected scenario of the CTI dependence on the shape of the 
signal packet (small CTI for a narrow shape as shown on the left-hand side and large CTI for a wider shape as shown on the right-hand side).}
\label{fig:diagram_cloudb}
\vspace*{-3mm}
\end{figure}

\begin{figure}[htp]
\centering
\includegraphics[width=\columnwidth,clip]{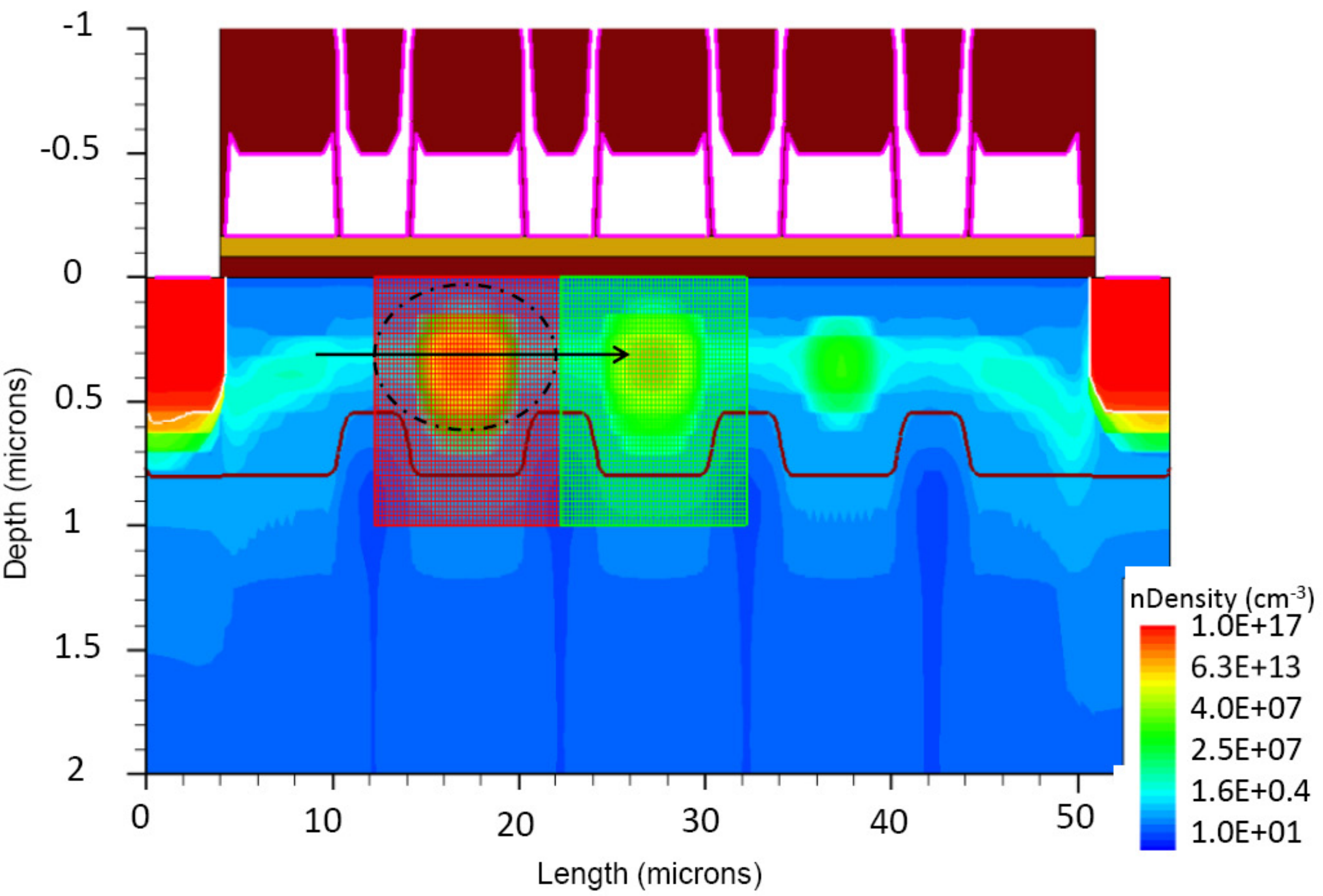}
\vspace*{-6mm}
\caption{Signal charge density in transit from a TCAD simulation of a CPCCD. The plot shows the charge packet 
located under a node at a depth of about $0.5$ microns. One pixel is located between $x=10$ and $30$ microns. The arrow 
indicates the direction of the transfer.}
\label{fig:charge_nodeb}
\end{figure}

\begin{figure}[htp]
\centering
\includegraphics[width=\columnwidth,clip]{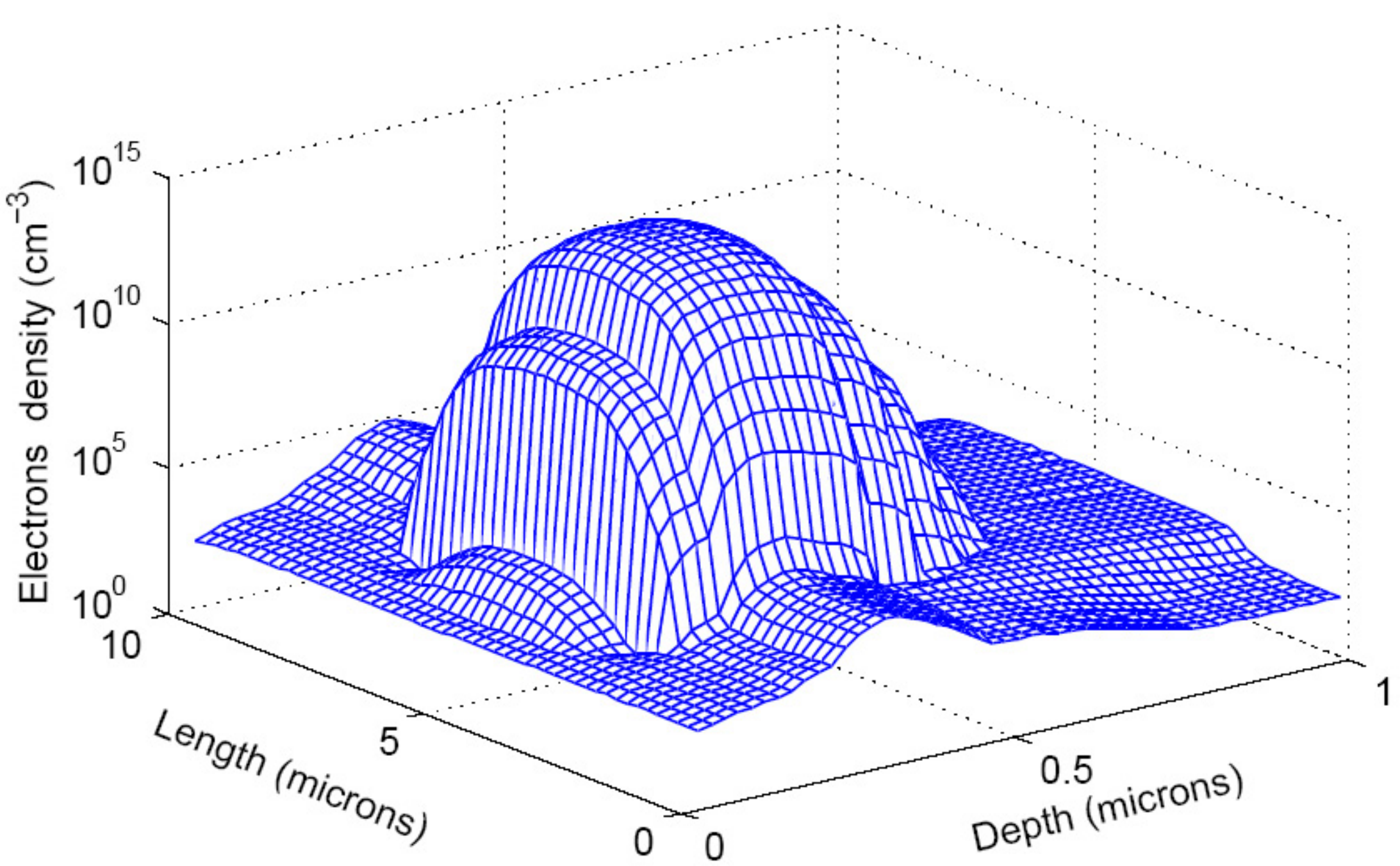}
\vspace*{-1mm}
\caption{Two-dimensional signal charge density extracted from the charge packet under one node using a full TCAD simulation.}
\label{fig:edensity2Db}
\end{figure}

\begin{figure}[htp]
\centering
\includegraphics[width=\columnwidth,clip]{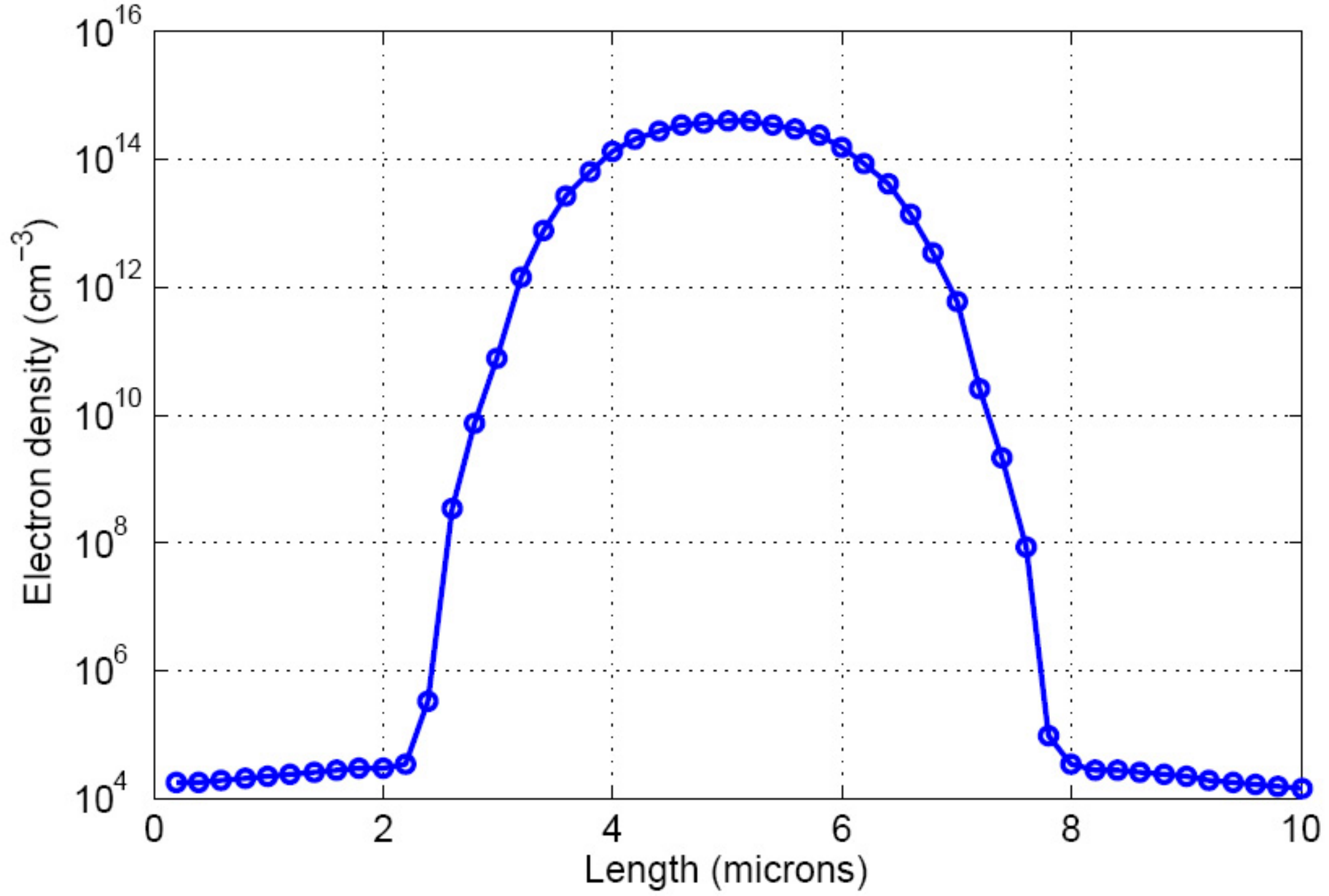}
\vspace*{-4mm}
\caption{One-dimensional signal charge density extracted from the charge packet under one node using a full TCAD 
simulation at a depth of $0.5$ microns.}
\label{fig:edensity1Db}
\end{figure}

The signal charge profile varies in the signal cloud as illustrated in the 
upper part of Fig.~\ref{fig:diagram_cloudb}. The signal packet 
does not have well defined boundaries and the charge concentration decreases gradually from 
the centre of the signal packet. Therefore, the signal packet will interact with a 
varying fraction of the traps within the pixel and this affects the CTI determination. The 
implementation of a  more realistic signal shape into the Analytic Model is expected to improve the agreement with the full TCAD simulation. 
Figure~\ref{fig:charge_nodeb} shows the profile of the signal charge under the node from a full TCAD simulation. Two-dimensional 
and one-dimensional signal charge density profiles are extracted as shown in 
Figs.~\ref{fig:edensity2Db} and~\ref{fig:edensity1Db}, respectively.
\vspace*{-3.5cm}
\vspace*{2cm}
\vspace*{1.5cm}

\begin{figure}
\centering
 \includegraphics[width=\columnwidth,clip]{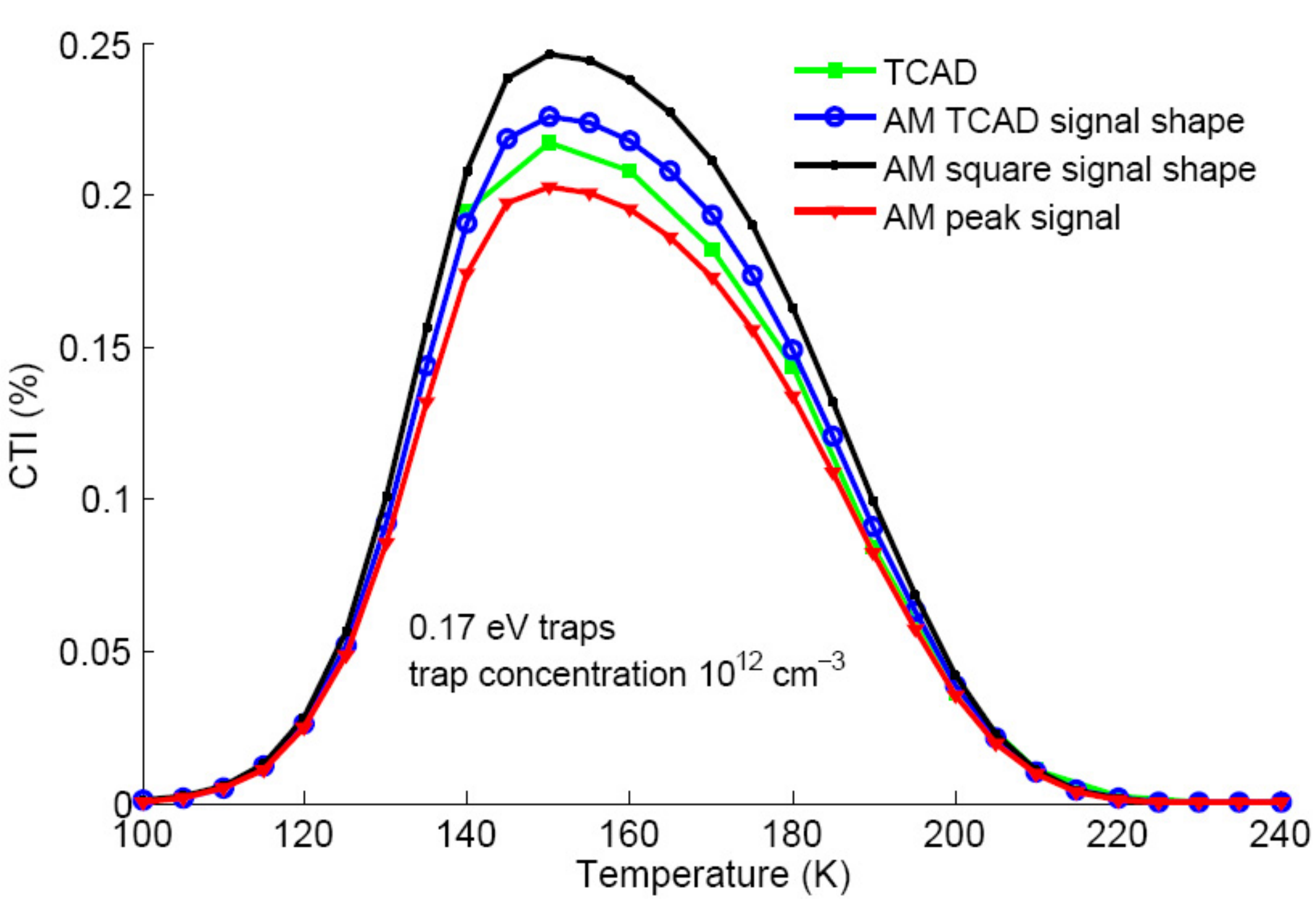}
\vspace*{-5mm}
\caption{CTI from Analytic Model (AM) including the shape of the signal packet as a function of temperature 
for 0.17 eV traps with a concentration of $10^{12}$ cm$^{-3}$ and $1 \%$ hit (pixel) occupancy at 50 MHz readout frequency in 
comparison with full TCAD simulation results. Three different signal shapes are compared with the full TCAD simulation. 
The CTI calculation with the Analytic Model using the signal shape extracted from a full TCAD simulation agrees better with the 
full TCAD simulation results than those from the Analytic Model using a square-shape signal as assumed previously.}
\label{fig:ctivstemp_signal17b}
\end{figure}

\begin{figure}
\centering
\includegraphics[width=\columnwidth,clip]{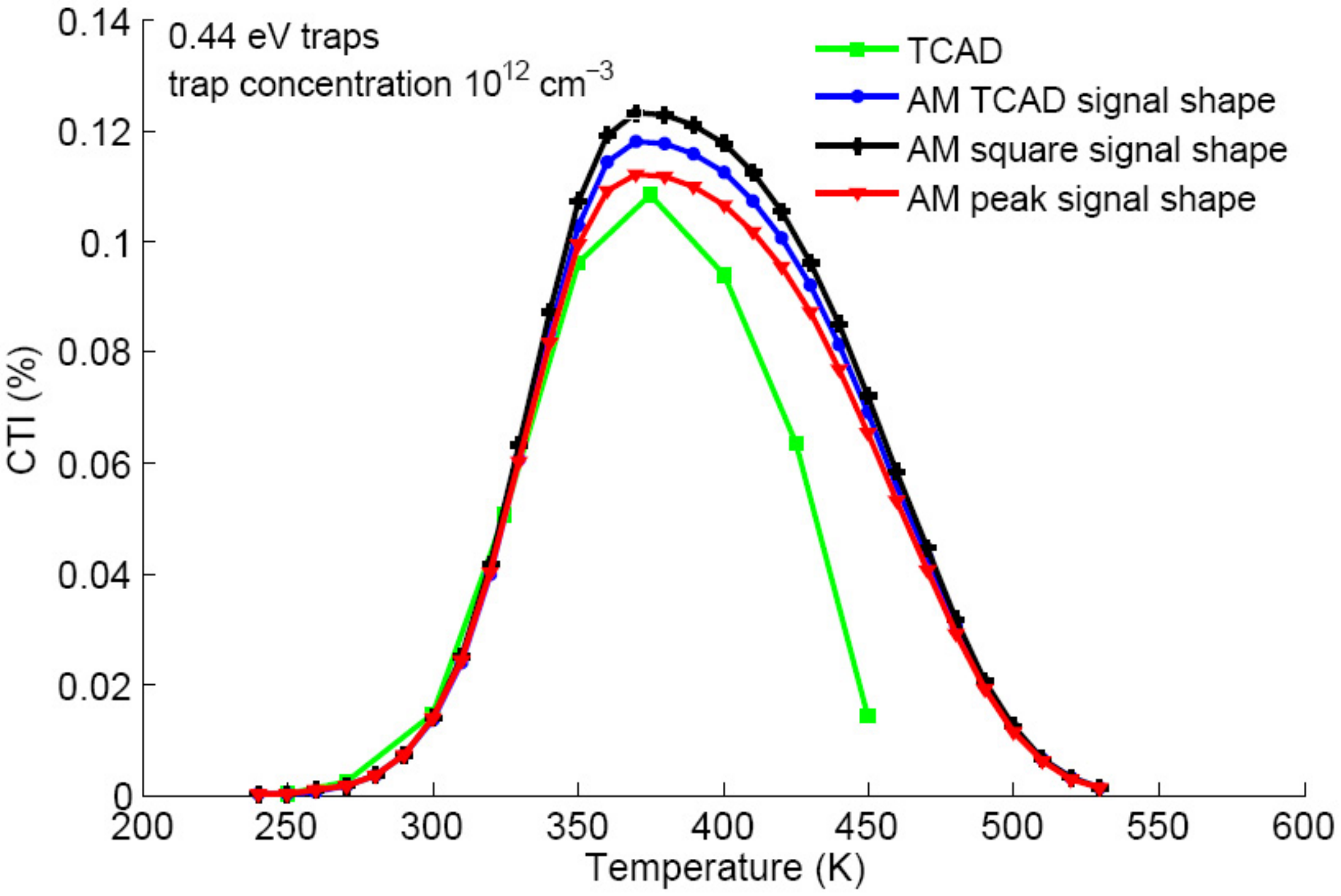}
\vspace*{-5mm}
\caption{CTI from  Analytic Model (AM) including the shape of the signal packet as a function of temperature 
for 0.44 eV traps with a concentration of $10^{12}$ cm$^{-3}$ and $1 \%$ hit (pixel) occupancy at 50 MHz 
readout frequency in comparison with full TCAD simulation results. Three different signal 
shapes are compared with the full TCAD simulation. For the 0.44 eV traps the inclusion of the signal shape in the 
Analytic Model has only a small effect to improve the agreement with the full TCAD simulation.}
\label{fig:ctivstemp_signal44b}
\end{figure}

Figures~\ref{fig:ctivstemp_signal17b} and~\ref{fig:ctivstemp_signal44b} show the CTI dependence on the signal charge profile 
for the 0.17 eV and 0.44 eV traps at 50 MHz. These figures also show the CTI values for different signal shapes in comparison 
with the full TCAD simulation. The CTI is reduced as the width of the potential well becomes smaller. This behaviour is expected,
as illustrated in the lower part of Fig.~\ref{fig:diagram_cloudb}. The CTI values calculated with the Analytic Model including 
the signal charge profile agree better with the full TCAD simulation results. The relatively shallow traps (0.17 eV) are more affected by the signal 
charge shape than the deeper ones (0.44 eV). The inclusion of the approximate signal shape in the Analytic Model reduces the CTI 
value in the peak region by about $10$ to $20 \%$ compared to assuming a square-shape signal.

\section{Comparison between full TCAD simulation and Analytic Model regarding clock voltage effect}
\begin{figure}
\vspace*{5mm}
\centering
\includegraphics[width=\columnwidth,clip]{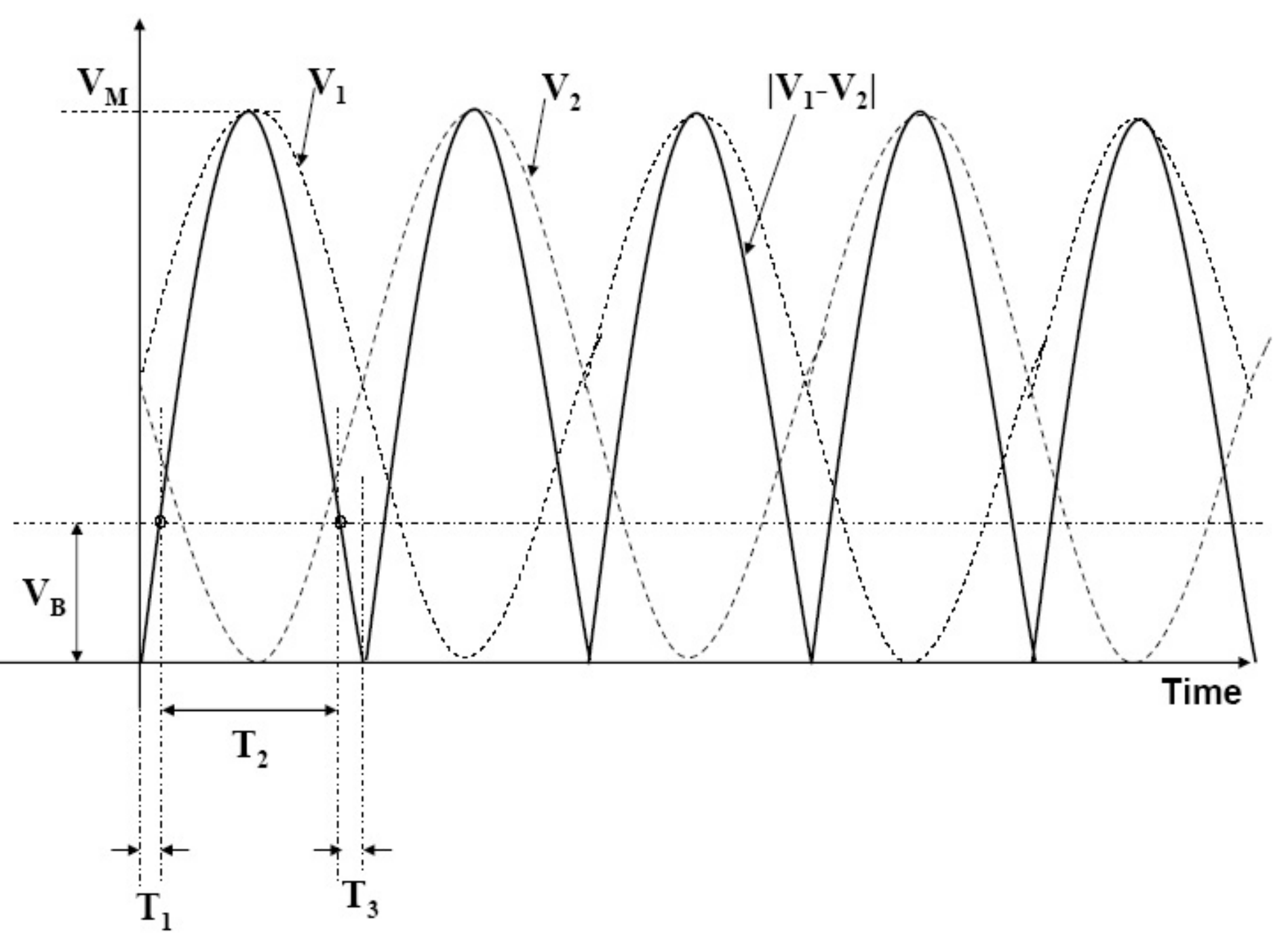}
\vspace*{-6mm}
\caption{Diagram of clock voltages in a two-phase CPCCD. $V_1$ and $V_2$ (dashed lines) show the applied voltages under node 1 and 2, 
respectively. The solid line shows the difference between the two applied voltages. $V_B$ is the barrier potential (horizontal dashed line). 
$T_1$ and $T_3$ are time periods with no charge transfer and $T_2$ is the time period with charge transfer.}
\label{fig:clockb}
\vspace*{5mm}
\end{figure}

\begin{figure}
\centering
\includegraphics[width=\columnwidth,clip]{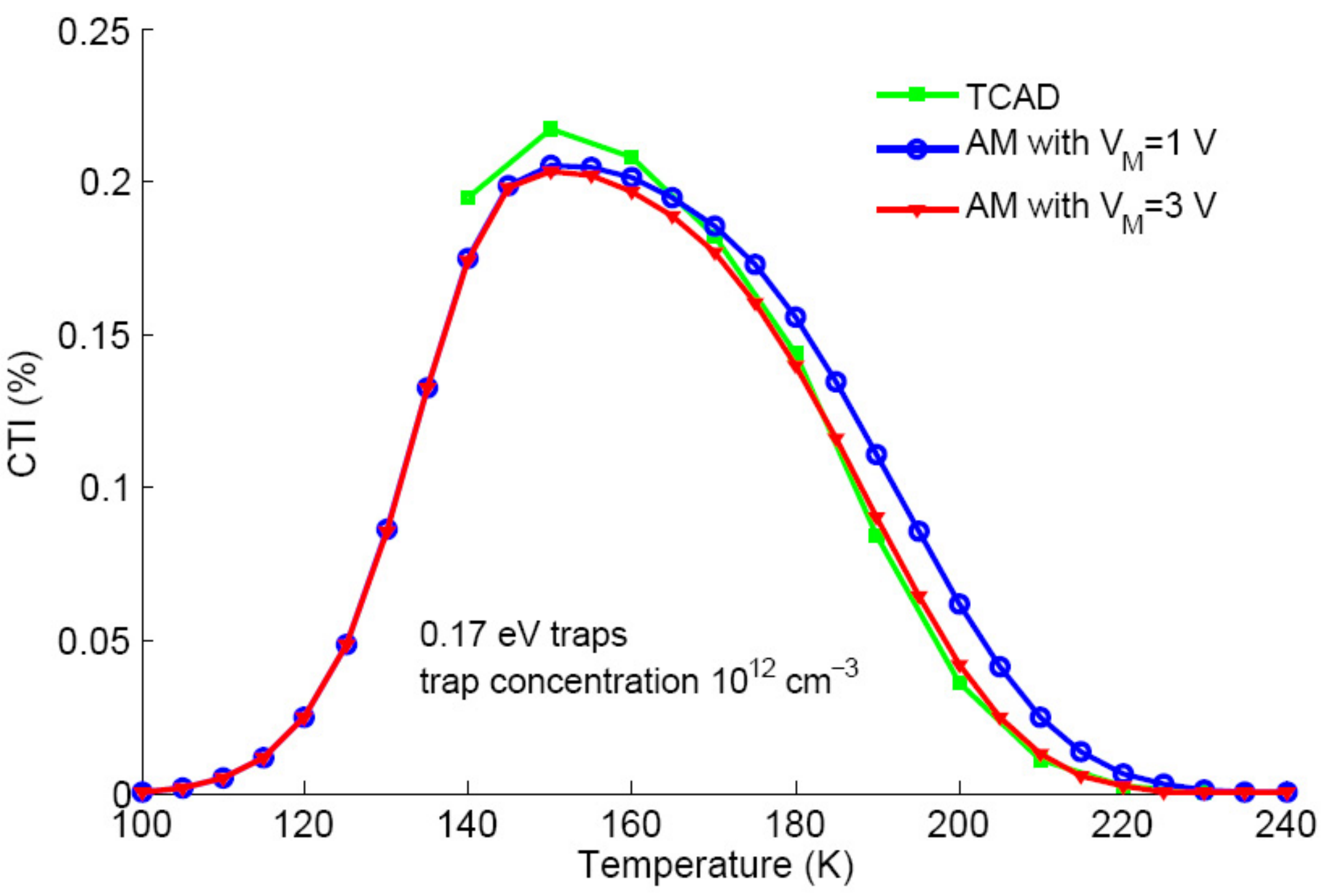}
\vspace*{-3mm}
\caption{CTI from Analytic Model (AM) including clock voltage effects as a function of temperature 
for 0.17 eV traps with a concentration of $10^{12}$ cm$^{-3}$ and $1\%$ hit (pixel) occupancy 
at 50 MHz readout frequency in comparison with full TCAD simulation results. Two different clock voltages ($V_M$) are shown.}
\label{fig:ctivstemp_clock17b}
\end{figure}

\begin{figure}
\centering
\includegraphics[width=\columnwidth,clip]{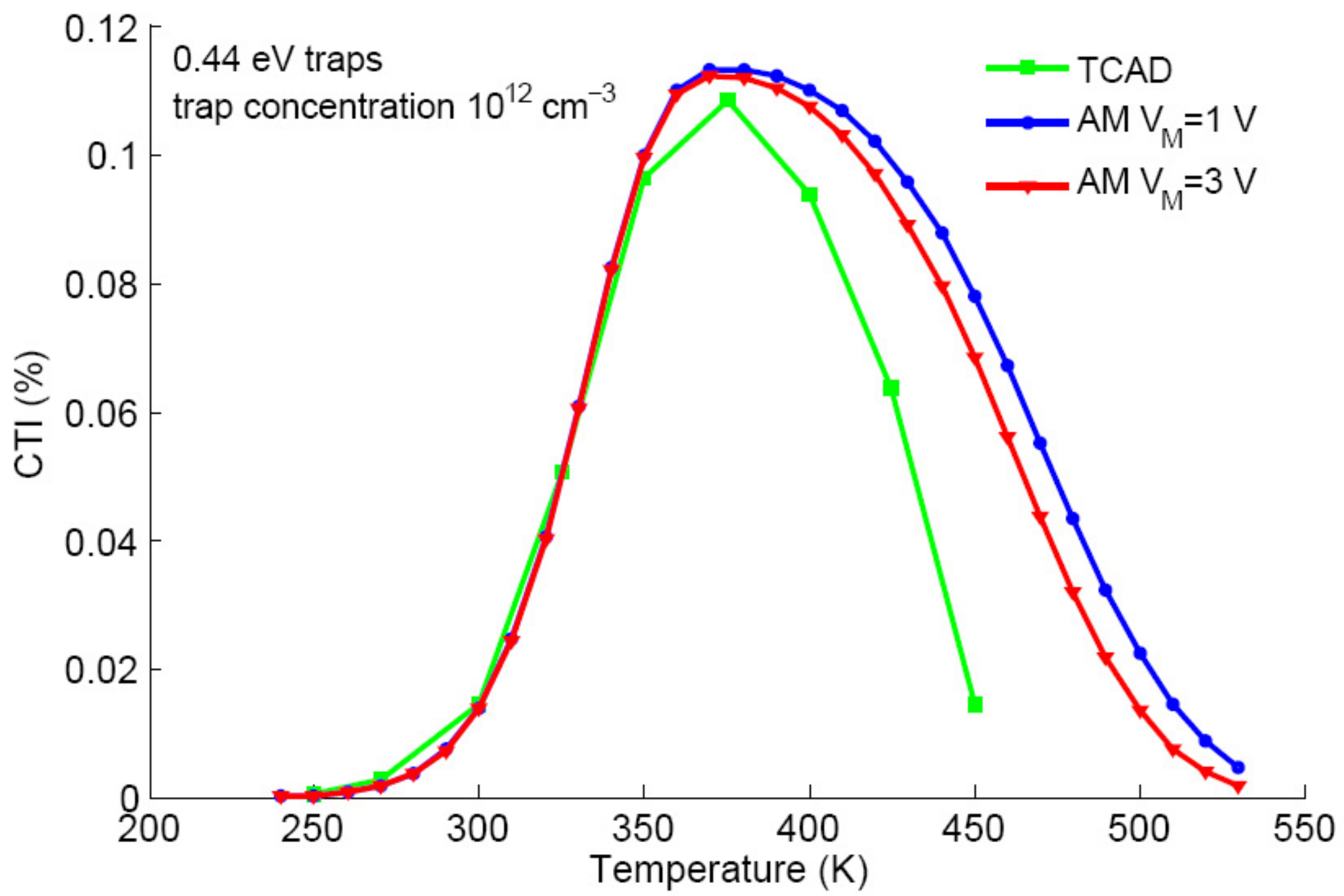}
\vspace*{-6mm}
\caption{CTI from Analytic Model (AM) including the clock voltage effects as a function of temperature 
for  0.44 eV traps with a concentration of $10^{12}$ cm$^{-3}$ and $1\%$ hit (pixel) occupancy 
at 50 MHz readout frequency in comparison with full TCAD simulation results. Two different clock voltages ($V_M$) are shown. }
\label{fig:ctivstemp_clock44b}
\end{figure}

In this study the effect of different clock voltage amplitudes on CTI values are investigated. A sine-form 
voltage is applied to consecutive nodes as shown in Fig.~\ref{fig:clockb}. The following variables are defined:
\vspace*{3cm}
\begin{itemize}
\item[$V_1$:] voltage applied to a first node of a pixel,
\item[$V_2$:] voltage applied to a second node of a pixel,
\item[$V_B$:] potential barrier created between two successive gates by the doping profile,
\item[$T_{1,3}$:] time interval where  $|V_1-V_2|$ $< V_B $,
\item[$T_2$:] time interval where  $|V_1-V_2|$ $> V_B $.\\
\end{itemize}
The signal is not transferred until the absolute difference between the two clock voltages $V_1$ and $V_2$ 
reaches the potential barrier created between two consecutive nodes. 
This affects the CTI determination and it is now included in the Analytic Model. The time intervals 
$T_1$ and $T_3$ are defined by the intersection point between the $|V_1-V_2|$ curve and the barrier 
potential $V_B$ (horizontal dashed line): $V_M\sin(wt)=V_B$, thus, 
\begin{equation}
T_1=T_2=\frac{1}{2\pi f}\times\sin^{-1}(\frac{V_M}{V_B})
\end{equation}
where $V_M$ is the amplitude of the clock voltage. The CTI determined with the Analytic Model including the clock voltage effect 
is shown in Figs.~\ref{fig:ctivstemp_clock17b} and~\ref{fig:ctivstemp_clock44b} for 0.17 eV and 0.44 eV traps, respectively. 
These results are compared to full TCAD simulations. Two different clock voltage amplitudes are used to 
illustrate the effect of the clock voltage. It is noted that the CTI decrease occurs only for temperatures above the CTI peak position.

\begin{figure}
\centering
\includegraphics[width=\columnwidth,clip]{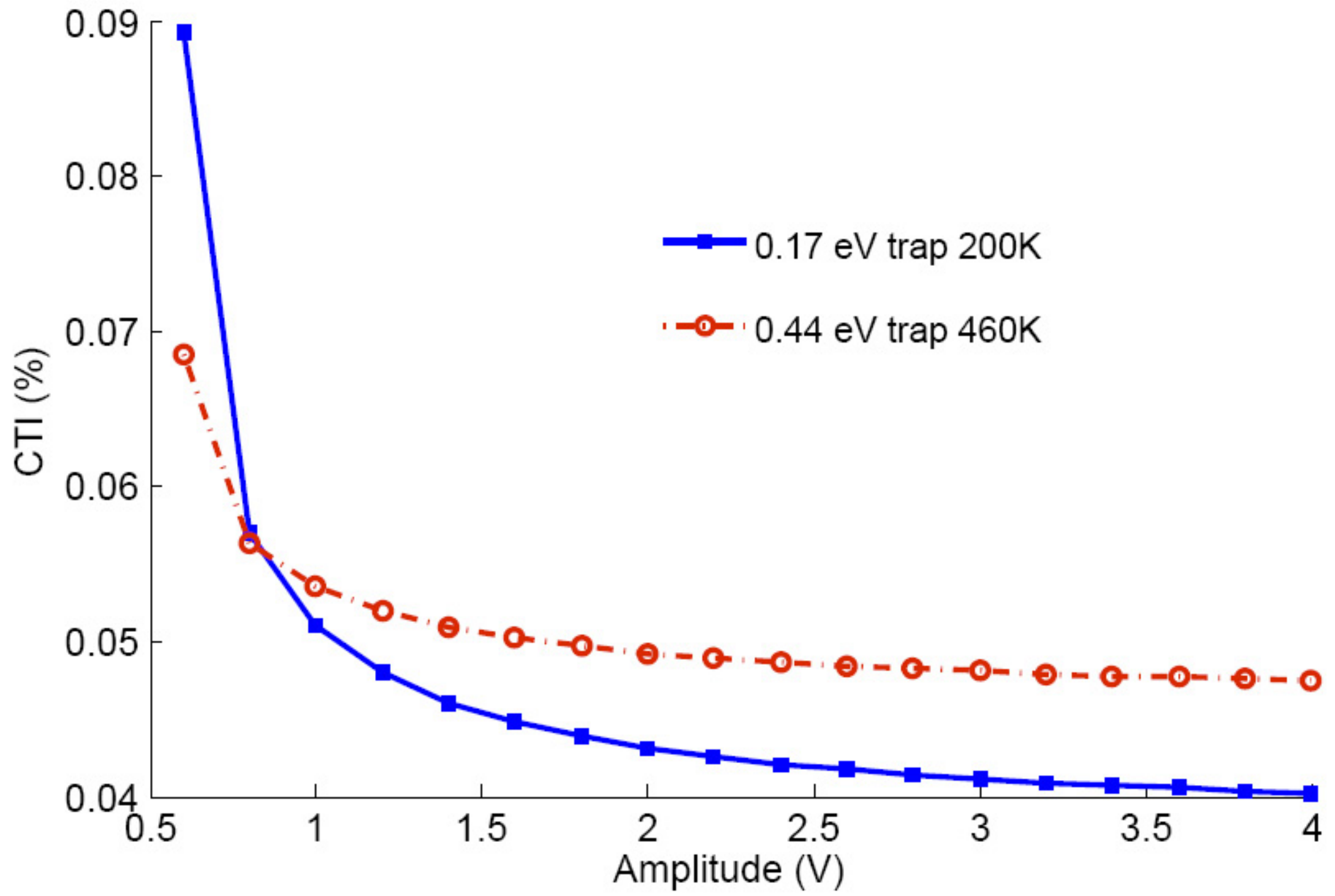}
\caption{CTI from Analytic Model as a function of clock voltage amplitude for both 
traps 0.17 eV and 0.44 eV with a concentration of $10^{12}$ cm$^{-3}$ at $T=200$ K and $T=460$ K, 
respectively, with $1\%$ hit (pixel) occupancy and 50 MHz readout frequency.}
\label{fig:CTIvsVM1744b}
\vspace*{-0.1cm}
\end{figure}

In addition, the effect of the clock voltage amplitude on the CTI is studied for a large variation of amplitudes. 
The CTI decreases as the amplitude increases until it saturates and no further decrease can be observed. 
This result is shown in Fig.~\ref{fig:CTIvsVM1744b} for two examples, 0.17 eV traps at a temperature of 200 K and 
0.44 eV traps at a temperature of 460 K.
\clearpage

\section{Conclusions and Outlook}
Our previous Analytic Models for a CPCCD have been extended to include the effect of non-uniform signal 
shape and the effect of realistic clock voltage amplitudes for CTI calculations. The signal shape 
affects the CTI mostly in the peak region. A smaller width of the potential well decreases the CTI. 
The inclusion of the clock voltage effects leads to smaller CTI values only above the CTI peak position. 
In summary, the Analytic Model has been extended to give a more realistic 
description of the CTI for a CPCCD and the results agree better with full TCAD simulations. Overall, 
the Analytic Model predicts well the CTI peak position in comparison with a full TCAD simulation. 
It can produce CTI values almost instantly while the full TCAD simulation is very CPU intensive. 
Generally, agreement between the Analytic Model and full TCAD simulation results is better for 
the 0.17 eV traps than for the 0.44 eV traps. The Analytic Model is suited to contribute to future CPCCD developments.

\section*{Acknowledgements}
\vspace*{-1mm}
This work is supported by the Science and Technology Facilities Council (STFC)
and Lancaster University. SA, KB and LD wish to thank the Algerian Government 
for financial support and Lancaster University for their hospitality. AS would like to thank the 
Faculty of Science and Technology at Lancaster University for financial support, and the organisers 
of the IEEE`08 and IPRD'08 conferences for their hospitality.

\end{document}